\documentclass[12pt]{article}
\usepackage{bbm} 
\usepackage{bm}

\usepackage{exscale}

\font\fourteenbf=cmbx12 scaled\magstep1

\def\a0size{6}

\newcommand{\lsi}{\raise0.3ex\hbox{$<$\kern-0.75em\raise-1.1ex\hbox{$\sim$}}}
\newcommand{\gsi}{\raise0.3ex\hbox{$>$\kern-0.75em\raise-1.1ex\hbox{$\sim$}}}
\newcommand{\lsim}{\mathop{\lsi}}
\newcommand{\gsim}{\mathop{\gsi}}
\renewcommand{\vec}[1]{{\bm #1}}

\begin{document}

\setlength{\baselineskip}{0.6cm}
\newcommand{\figysize}{16.0cm}
\newcommand{\figtopspace}{\vspace*{-1.5cm}}
\newcommand{\figbottomspace}{\vspace*{-5.0cm}}
  

\begin{titlepage}
\begin{flushright}
BI-TP 2005/30
\\
\end{flushright}
\begin{centering}
\vfill

{\fourteenbf \centerline{ 
The impact of QCD plasma instabilities 
}
\vskip 2mm \centerline{ on bottom-up thermalization 
}}

\vspace{1cm}

Dietrich B\"odeker \footnote{e-mail: bodeker@physik.uni-bielefeld.de}

\vspace{.6cm} { \em 
Fakult\"at f\"ur Physik, Universit\"at Bielefeld, 33615 Bielefeld, Germany
}

\vspace{2cm}
 
{\bf Abstract}

\end{centering}
 
\vspace{0.5cm}

QCD plasma instabilities, caused by an anisotropic momentum distributions
of the particles in the plasma,
are likely to play an important role in  thermalization
in heavy ion collisions. We consider plasmas with two different components 
of particles, one  strongly  
anisotropic and one isotropic or nearly isotropic. The isotropic component
does not eliminate instabilities but it decreases their growth rates. 
We investigate the impact of plasma instabilities
on the first stage of the ``bottom-up'' thermalization scenario in which
such a two-component plasma emerges, and find
that even in the case of non-abelian saturation instabilities
qualitatively change the bottom-up picture. 

\noindent

\vspace{0.5cm}\noindent

 
\vspace{0.3cm}\noindent
 
\vfill \vfill
\noindent
 
\end{titlepage}
 
\section{Introduction}
\label{sc:introduction} 

It is  an interesting and still open
question whether in heavy ion collisions it is possible
to create a medium made of quarks and gluons which is locally in thermal
equilibrium, which would 
allow to study thermodynamics of QCD. On the theoretical side it
would be desirable to have an answer to this question in a limit when
one has perturbative control, which could be the case for very large collision
energies. 

This question can probably be divided into two. The first concerns the earliest
stage of the collision when partons are ``freed'' from the colliding nuclei. 
The second is how this system subsequently evolves in time, and
whether and when it approaches local thermal equilibrium. It should
be possible to address the second question 
in perturbation theory, provided the produced partons
have sufficiently large energies and are sufficiently dense so that multiple
scatterings occur. 

In the saturation scenario for high energy nucleus-nucleus scattering
\cite{saturation} the freed partons are mostly gluons   
with transverse momenta of order of the so called saturation scale 
$ Q \gg \Lambda  _{\rm QCD}$. 
The distribution function in phase space, or
occupation number  $ f ( x, \vec p )  $  
of these gluons is of order $ 1/g ^ 2  $, where $ g $ is the QCD gauge coupling
($ \alpha  _ {\rm s} = g ^ 2/( 4 \pi   ) $). The scale that sets the value of  
$ g $ is expected to be $ Q $, so that for large collision energies the 
occupation number is large. Since the typical occupation number in thermal
equilibrium is of order one, this system is far from equilibrium. 

It is customary
to assume that $ f $ does not depend on the coordinates transverse  to the beam
direction, which we choose as the $ z $-axis, and that $ f $ is 
invariant under rotations around $ \hat{\vec z} $. This should be a good
approximation for sufficiently large nuclei and for sufficiently central
(head-on) collisions. 
Furthermore, it is assumed that the gluon distribution function is invariant under 
boosts in $ z $-direction. With these assumptions $ f $ only depends on  
$ \tau   = \sqrt{t ^ 2 - z ^ 2 }$, on the magnitude of the momentum
perpendicular to $ \hat{\vec z} $, and on the difference 
of the momentum- and coordinate-space rapidities. 

After the gluons are freed around $ Q \tau  \sim 1 $, 
the system expands and the occupation number of the
originally 
produced, so called ``hard'' gluons drops below $ 1/g ^ 2 $. 
Without 
production of additional gluons their number density $ n _ {\rm h} $ decreases like 
$ 1/\tau   $. It is not at all obvious \cite{muellerBoltzmann,raju}  
that such a system thermalizes which 
would require that the relevant interaction rate is
larger than the expansion rate $ 1/\tau  $. Consider for  instance the rate $
\Gamma  $ for $ 2 \to 2 $ large angle scattering of hard gluons for which the cross 
section $ \sigma  $ is of order $ g ^ 4/Q ^ 2 $. Taking into account Bose
enhancement and using 
$ n _ {\rm h} \sim n _{\rm h, initial} 
/ ( Q \tau) \sim Q ^ 3/( g ^ 2 Q \tau) $ one can estimate 
\begin{eqnarray} 
    \Gamma   \sim \sigma  n _ {\rm h}( 1 +  f _ {\rm h} )  
    \sim g ^ 2 ( 1 + f _ {\rm h}) 
    /\tau   
\end{eqnarray} 
which is always small compared to the expansion rate.

In Ref.~\cite{wong} it was argued that inelastic processes play an essential
role for thermalization. In the detailed thermalization 
scenario of Ref.~\cite{bottomUp} ``soft'' gluons are produced by
bremsstrahlung off the hard gluons. Initially most of the soft gluons
have the smallest 
possible energy, but then their momentum rapidly increases in multiple elastic
scatterings with the hard gluons. The soft gluons thermalize among themselves,
and are then  heated by the hard gluons which eventually loose all their
energy to the soft-gluon heat bath (``bottom-up thermalization'').

While there is no minimal momentum transfer for massless gluon exchange in the
vacuum, a  
medium usually cuts off long range interactions. For particles with nearly 
isotropic momentum distributions this can be accounted
for by including the polarization tensor in the propagators of the exchanged
gluons 
\footnote{The situation is more complicated for 
magneto-static interactions, but at
  leading order these do
  not play a role in the scattering processes which are relevant to
  thermalization.}. 
However, due to the longitudinal expansion the typical longitudinal momentum
of hard gluons $ p _ {\rm h} ^ z $ is much smaller than the transverse momentum
which for a long time remains order $ Q $, so that the momentum distribution
of the hard gluons is strongly anisotropic. 
Anisotropic distributions can cause so called plasma
\footnote{Here ``plasma'' refers to a system of quarks and gluons
which is not necessarily in thermal equilibrium, while sometimes the term 
``quark-gluon-plasma'' is used only for  thermalized or almost thermalized systems.} 
instabilities which means that 
some long wavelength ($  | \vec k | \ll Q $) gauge field modes 
grow exponentially if their amplitudes are sufficiently small. This is a
collective phenomenon
which is not visible in the kinetic equation framework
of this problem used in 
\cite{muellerBoltzmann,raju,wong,bottomUp}. 
The effect has been known for many years in
plasma physics,  and it has been argued  \cite{mrowczynskiInitial}
that QCD plasma instabilities 
can speed up equilibration in heavy ion collisions since they tend to make
the momentum distributions more isotropic.   

The possible relevance of plasma instabilities to the bottom-up scenario was
realized in Ref.~\cite{arnoldInstabilities} where also the qualitative
difference between QED and 
QCD plasma instabilities was discussed.
At some point the growth of instabilities is stopped by non-linear effects. In
QED this happens when the amplitude of the unstable modes has become so large
that they deflect a particle's momenta by a large angle  over
the distance of one wavelength of an unstable mode. 
This corresponds to gauge field amplitudes 
$ A \sim p / e $ where $ p $ is a typical particle momentum. Then plasma
instabilities have a dramatic effect, in particular they are able to
isotropize the particle's momentum distribution within a  very short time.
In QCD gluons are self-interacting which may change the behavior of
instabilities completely. The linear approximation already breaks down at much
smaller 
amplitudes $ A \sim k/g $ where $ |  \vec k  |  \ll |  \vec p |  $ 
is a characteristic wave vector of
$ A $, and a crucial question is  whether these non-linearities stop the
growth of instabilities. In Ref.~\cite{arnoldAbelianization} it was suggested 
that gluon self-interactions may not saturate the  
instabilities because the system can ``abelianize''  so that 
the unstable modes can grow until they hit the abelian saturation
bound $ A \lsim p/g $. In this case the distribution of 
hard gluons would quickly become isotropic.
It has been argued in Ref.~\cite{arnoldThermalization} that this is 
sufficient for 
a hydrodynamic description to be applicable even if there is no local thermal
equilibrium. This could solve the puzzle 
\cite{hydrodynamics} why hydrodynamic
calculations successfully describe experimental results for elliptic flow
provided that  
they can be used from very early times ($ \tau  \sim 0.6 $ fm$ /c $) on,  while 
perturbative estimates for thermalization times are substantially larger 
($ \tau  \gsim  2.5 $ fm$/c $) \cite{arnoldThermalization}. 
In lattice simulations with fields only depending on $ t $ and $ z $
abelianization was indeed observed \cite{romatschke2d,dumitru}, but the
recent 3+1 dimensional simulations of Refs.~\cite{arnoldFate,romatschkeFate} indicate
that instabilities  are saturated  
by non-abelian interactions which would mean  that their  effect is  less
dramatic than suggested in Ref.~\cite{arnoldThermalization}. Nevertheless 
the instabilities  lead to larger infrared gluon fields 
than in an isotropic plasma 
and their role in the evolution of the system remains to be understood.  

So far only the effect of the hard, strongly anisotropic gluons on the infrared
modes has been considered in the literature. In the bottom-up scenario also the
produced soft gluons, which have isotropic or nearly isotropic momentum
distribution,  contribute to the polarization tensor and can therefore
have an influence on plasma instabilities as recently emphasized in 
Ref.~\cite{muellerModified}. In Sec.~\ref{sc:two} we consider 
plasma instabilities in a system containing two components, one strongly
anisotropic and one isotropic. In Sec.~\ref{sc:bottomUp} we show how in the early
stage of the bottom-up scenario plasma instabilities affect the momentum
distribution of hard gluons even in the case of  non-abelian saturation.

{\it Notation}: 4-vectors are denoted by lower case italics, 
3-vectors by boldface. The
metric is ``mostly negative'',  $ ( g _  {\mu  \nu  } ) = \mbox{diag}(1, -1,
-1, -1) $. 

\section{Instabilities in two-component plasmas}
\label{sc:two} 

In the presence of instabilities one cannot naively resum the polarization
tensor into the gluon propagator and then use it to compute scattering 
matrix elements entering a Boltzmann equation. 
Instead, one has to fully take into account the dynamics
of the 
low momentum gauge  field modes. Typically they have 
large occupation numbers and can be described by a classical
gluon field $ A _ \mu  = A _  \mu  ^ a T ^ a $, where $ T ^ a $ are the
SU(3) generators in the fundamental representation. 
When the occupation numbers of high momentum gluons are small compared to 
$  1/g ^ 2 $  
they can be described as weakly interacting particles with a definite
momentum. The dynamics of this coupled system  satisfies the
non-abelian Vlasov equations  \cite{heinz,vlasovQED}
\begin{eqnarray}
  ( D _  \mu    F ^  {\mu  \nu  } ) ^ a 
  = g \int \frac{ d ^ 3 p }{(2 \pi ) ^ 3} v ^ \nu  
  f ^ a,
  \label{maxwell} 
\\
   v \cdot D \hat f 
   + \frac12 g v ^ \mu \bigg \{  F _{\mu  i } , \frac{ \partial \hat f }{\partial 
    p ^ i} \bigg \} = 0
  \label{vlasov} 
\end{eqnarray} 
where $ D _ \mu  = \partial _ \mu  - i g [A _ \mu , \cdot]$ 
is the adjoint representation 
covariant derivative,  $ F ^{\mu  \nu  } $ is the field strength tensor 
and  $ v ^ \mu  \equiv (1,  \hat{\vec p}  ) $. The distribution functions
of high momentum gluons are encoded in $ \hat f ( x, \vec p ) $ which 
is a hermitian $ N \times N $ matrix with $ N = 3 $ being
the number of colors and $ f ^ a = {\rm tr} ( T ^ a \hat f  ) $. 
For sufficiently small 
gauge field amplitudes one can write 
$ \hat f =   f  \mathbbm{1}/N + \delta  \hat f $ 
and linearize with respect to $ A $ and  $ \delta  \hat f $. 
In order to see whether there are instabilities one can neglect the 
$ x $-dependence of 
$ f $ provided the growth rate and wave vectors  of the unstable modes are
large compared to the inverse of the time and lenght scales on which $ f $ varies. 
The initial value problem can then be solved by spatial Fourier 
transformation and one-sided  Fourier transformation with respect to time, 
$ A ( k _ 0 ) \equiv  \int _ 0 ^ {\infty  } d t e ^{i k _ 0 t } A ( t )   $ 
where the frequency $ k _ 0 $
is in the upper half of the complex plane \cite{landau10}\footnote{This is 
of course nothing but a Laplace transformation with a different
  convention  for the variable conjugate to $ t $.}.
 By eliminating $ \delta  \hat f ( k ) $ one obtains an equation for 
$ A ( k ) $ which takes the form 
\begin{eqnarray}
  \Big [ k ^ 2 g ^{\mu  \nu  } - k ^ \mu  k ^ \nu  
    + \Pi  ^{\mu  \nu  } ( k ) \Big ] A _ \nu  ( k ) = \Phi  ^ \mu 
  ( k )
  \label{phi} 
\end{eqnarray} 
with the polarization tensor
\begin{eqnarray}
  \Pi  ^{\mu  \nu  }    ( k ) 
   &=&  2 N  g ^ 2 
   \int \frac{ d ^ 3 p }{(2 \pi ) ^ 3}\frac{ \partial f ( \vec p ) }{\partial p ^ i}
   \left ( v ^ \mu  \delta  ^{i \nu  } - \frac{ v ^ \mu  v ^ \nu k ^ i }{v \cdot k}
     \right )
    \nonumber\\
   &=& 2 N  g ^ 2 
   \int \frac{ d ^ 3 p }{(2 \pi ) ^ 3}\frac{  f ( \vec p ) }{|\vec p |}
   \left ( - g ^{\mu  \nu  } 
   + \frac{k ^ \mu  v ^ \nu  + k ^ \nu  v ^ \mu }{v \cdot k}
     - \frac{ v ^ \mu  v ^ \nu  k ^ 2 }{( v \cdot k ) ^ 2}
     \right )
 \label{polarization} 
\end{eqnarray}  
In Eq.~(\ref{phi}) 
$ \Phi  ^ \mu  $ is an analytic function of $ k ^ 0 $ off the real axis which
depends linearly on  the initial values at $ t = 0 $
for $ A  ( \vec k ) $, $ \dot A ( \vec k ) $, and $ \delta  \hat f (
\vec k, \vec p) $. 
After fixing the gauge one can solve Eq.~(\ref{phi}) for $ A ( k ) $.  
If $ f  ( \vec p ) $ depends on the direction of $ \vec p $, $ A ( k ) $ in general
has poles on the imaginary $
k ^ 0 $-axis which yield exponentially growing solutions for $ A ( t, \vec k
)$.  

Following Ref.~\cite{arnoldInstabilities} we introduce
 \footnote{In \cite{arnoldInstabilities} 
$ m ^ 2 $ is called $ m ^ 2 _{{\infty  } } $, and parametrically it is the same as
 $ m^2_{\rm D} $ of Ref.~\cite{bottomUp}. }
\begin{eqnarray}
  m ^ 2 \equiv 2 N  g ^ 2\int \frac{ d ^ 3 p }{(2 \pi ) ^ 3} 
  \frac{ f ( \vec p ) }{|\vec p |}
  \label{m2} 
\end{eqnarray} 
in order to characterize the size of $ \Pi ^ {{\mu \nu}}  $. The 
physical interpretation of $ m ^
2 $ depends on further properties of $ \Pi  ^{\mu  \nu  }$.  
For example, in an isotropic plasma  
gluon fields with $ k _ 0 \sim |\vec k |
\lsim  m  $ 
are dynamically screened. For anisotropic plasmas there are
unstable gauge field modes which grow exponentially with 
a growth rate of order $  m  $ or smaller. For a mildly anisotropic plasma
only modes with $  |   \vec k  |   \lsim  m $ are unstable while for strongly
anisotropic plasmas there are also unstable modes with $  |   \vec k  |   \gg
m $. 

In the bottom-up scenario there are contributions to the polarization
tensor from different scales for $ \vec p $. For example, in 
the first stage ($ 1 \ll Q \tau  \ll g ^{-3} $)
$ m ^ 2 $ receives  contributions from hard and soft 
gluons which are parametrically of the same size. 
In this case $ m ^ 2 $ is the sum of $  m ^ 2 _ {\rm h} $ and $  m ^ 2 _ {\rm s} $, which  
are defined as in Eq.~(\ref{m2}), 
but with the $ \vec p $-integration restricted to hard and soft momenta,
respectively.  
The hard gluons
are strongly anisotropic, while the soft gluons are close to 
being isotropic. Therefore there might be no instabilities 
if  only the soft gluons were present. 
For the subsequent discussion the magnitude of the
hard and soft momenta play no role. These only enter the values of 
$ m _ {\rm h} $ and $ m _ {\rm s} $, so we can simply think of the high
momentum  gluons as some two-component plasma.

Naively one might think that if one adds some isotropically distributed
particles the instabilities might disappear.  That this is not necessarily
the case can be easily understood from the criterion for the occurrence
of instabilities of Ref.~\cite{arnoldInstabilities}.  It 
states that there
is an instability associated with a given wave vector $ \vec k $ for each 
negative eigenvalue of the matrix $ \lim _{\epsilon  \to 0} [ \vec k ^ 2 \delta  ^{ij} 
- k ^ i k ^ j + \Pi  ^{ij} ( i \epsilon  , \vec k ) ]$. 
For an isotropic distribution 
$ \Pi  ^{ij}( k )  $ vanishes in this limit,  so  adding
particles with isotropic distribution does not affect the occurrence of plasma
instabilities 
\footnote{
It is of course possible that the added particles increase the collision 
rate of the gluons which  contribute to $ f $ to the extent that the collisionless
approximation, which was used to obtain this criterion, breaks down. Then
the instabilities might disappear.
}. 

We write $ f = f _ {\rm h} +  f _ {\rm s} $ and correspondingly $ 
  \Pi ^ {{\mu \nu}}  
= \Pi ^ {{\mu \nu}}   _ {\rm h} + \Pi ^ {{\mu \nu}} _ {\rm s} $ 
and we assume that $ f _ {\rm s}  $ only depends on $   |     \vec p   |    $. Then 
$ \Pi ^ {{\mu \nu}}    _ {\rm s} $ has exactly the same form as a polarization   
tensor in thermal equilibrium \cite{kraemmer} with the Debye mass 
$ m_{\rm D}   $ replaced by $ \sqrt{2} \; m _ {\rm s}$.

We are interested in hard gluon distributions for which the typical
$ p _ z $ is small compared to its transverse momentum, i.e., the typical 
$ z $-component of $ \vec v $ in Eqs.~(\ref{polarization}) $    v _ {\rm h} ^ z  $ is 
much smaller than one. Detailed discussions of the resulting polarization
tensor can be found in Refs.~\cite{arnoldInstabilities,romatschkePi}. 
There are no instabilities for $ |  \vec k |  \gg  k _{\rm max} $ 
with 
\begin{eqnarray}
    k _{\rm max} \sim \frac{ m _ {\rm h} }{v _ {\rm h} ^ z }
    \label{kmax} 
\end{eqnarray} 
To keep the present discussion as simple as possible we consider 
$  |   \vec k _\perp  |
\ll  |   k _ 0  |   $ and $  v _ {\rm h} ^ z |   k _ z  |   \ll  |   k _ 0  | $ 
so that we can 
approximate $ v \cdot k \simeq k _ 0 $ in the second line of 
Eq.~(\ref{polarization}). We
assume that $ f _ {\rm h} $ is invariant under $ \vec p \to -\vec p $ and under
rotations around the $ z $-axis. Then one finds 
\begin{eqnarray}
  \Pi  _ {\rm h} ^{ij} ( k ) \simeq \left ( \delta  ^{ij} - \delta  ^{iz}  
  \delta  ^{jz} \right ) \Pi  _ {\rm h} ^{{} ^\perp } (  k ) 
  \label{impact1.2a} 
\end{eqnarray} 
with 
\begin{eqnarray}
  \Pi   _ {\rm h}  ^{{} ^ \perp} ( k ) 
   =  \frac{ m _ {\rm h} ^ 2}{2} \frac{ k _ 0 ^ 2 + \vec k ^ 2
  }{k _ 0 ^ 2} 
  \label{impact1.2b} 
\end{eqnarray} 
Note that this is negative when $ k _ 0 = i \gamma  $ and 
$ \gamma  ^ 2 < \vec k ^ 2 $. 
With the same approximations the soft contribution to $ \Pi  ^{ij} $ 
can be written as  
\begin{eqnarray}
  \Pi  _ {\rm s} ^{ij} ( k )  \simeq 
  \left ( \delta  ^{ij} - \delta  ^{iz} \delta  ^{jz} \right ) 
  \Pi  _ {\rm s} ^{{} ^\perp }    ( k )  + 
  \delta  ^{iz} \delta  ^{jz} \Pi  _ {\rm s} ^{zz }    ( k )
\end{eqnarray} 
Then there are two unstable modes with polarization approximately orthogonal
to $ \hat{\vec z}  $ and their growth rate $ \gamma  $ is determined by
\begin{eqnarray}
  \gamma  ^ 2 + \vec k ^ 2 + \Pi  ^ {{} ^\perp} ( i \gamma , \vec k ) = 0
  \label{equation} 
\end{eqnarray} 
where $ \Pi  ^ {{} ^\perp} = \Pi  ^ {{} ^\perp}  _ {\rm h}
+ \Pi  ^ {{} ^\perp}  _ {\rm s}$.  
As we have discussed above, including $ \Pi  ^{{}   ^\perp} _ {\rm s} $ 
in Eq.~(\ref{equation}) does not eliminate instabilities but it decreases
their growth rate because both 
$  \Pi  ^{{} ^\perp} _ {\rm h} ( i \gamma , \vec k ) $ and 
$  \Pi  ^{{} ^\perp} _ {\rm s}
( i \gamma , \vec k )  $ increase monotonically with increasing $ \gamma   $
and because $  \Pi  ^{{}   ^\perp} _ {\rm s} ( i \gamma , \vec k ) \ge 0$.

First consider the case $ m _ {\rm s} \sim m _ {\rm h} $. 
For $ |\vec k|  \sim m _ {\rm h}  $ there
is only one scale in Eq.~(\ref{equation}) 
and the instabilities grow with a rate of order $ m _ {\rm h} $.
When $  k _ {\rm max} \gg |  \vec k  |  \gg m _ {\rm h} $, 
one would have  $ \gamma  \simeq m _ {\rm h} /\sqrt{2}$ if the soft gluons were
absent.  Since the soft gluons decrease the
growth rate we  have $ \gamma  \ll  | \vec k  |  $.  
Then one can approximate
\begin{eqnarray}
  \Pi  ^{{} ^\perp}  _ {\rm s} ( k ) \simeq 
   - i \frac{ \pi }{2} m _ {\rm s} ^ 2 \frac{ k _ 0}{|\vec k|}
  \qquad (|k _ 0| \ll |\vec k|, \quad  {\rm Im}( k ^ 0) > 0)
   \label{impact.9} 
\end{eqnarray} 
Using Eqs.~(\ref{impact1.2b}) and (\ref{impact.9}) and neglecting
$ \gamma   ^ 2 \ll \vec k ^ 2 $,  Eq.~(\ref{equation}) becomes
a cubic equation for $ \gamma  $,
\begin{eqnarray}
  \vec k ^ 2 - \frac12 m _ {\rm h} ^ 2 \frac{  \vec k ^ 2}{ \gamma  ^ 2}   + \frac{ \pi
  }{2} m _ {\rm s} ^ 2 
  \frac{ \gamma  }{|\vec k|} = 0
   \label{cubic} 
\end{eqnarray} 
Since $ \gamma  \ll  |  \vec k  |  $ the third term on the left hand side is small
compared to the second and can be neglected which again gives 
\begin{eqnarray} 
  \gamma  \simeq \frac{ m _ {\rm h}}{\sqrt{2}} 
  \qquad ( m _ {\rm h} \sim m _ {\rm s}, \; m _ {\rm h}
  \ll |  \vec k |  \ll k _{\rm max} ) 
  ,
\end{eqnarray} 
i.e., in this regime the effect of  soft gluons on the growth rate is negligible. 

Now we consider what happens if $ m _ {\rm s} \gg m _ {\rm h} $.  We will find growth
rates which are small compared to $ |\vec k |  $. Therefore we can use
Eq.~(\ref{cubic}) to compute $ \gamma   $.
First consider small $ |  \vec  k |  $. We neglect the first term in Eq.~(\ref{cubic})
which will be justified  in a moment to obtain
\begin{eqnarray}
  \gamma  \simeq   \left ( \frac{ m _ {\rm h} ^ 2}{ \pi  m _ {\rm s} ^ 2} \right ) ^{1/3} | \vec k|
\end{eqnarray} 
Now we see that one can indeed neglect the first term in Eq.~(\ref{cubic}) 
as long as $ | \vec k| \ll ( m _ {\rm s} / m _ {\rm h} )^{2/3} m _ {\rm h} $. 
In this range the growth rate is reduced by the soft gluons and it
increases linearly with $  |  \vec k  | $. Going to larger $  |  \vec k  |  $
 all terms in Eq.~(\ref{cubic}) become of the same order of magnitude and the
 full cubic 
 equation for $ \gamma  $ needs to be solved. Finally, when 
$ | \vec k| \gg ( m _ {\rm s} / m _ {\rm h} )^{2/3} m _ {\rm h}$ 
one can neglect the last term on
the left hand side of (\ref{cubic}) which gives
\begin{eqnarray}
  \gamma  \simeq \frac{ m _ {\rm h}}{\sqrt{2}  }  
  \qquad ( m _ {\rm h} \ll  m _ {\rm s}, \;
  m _ {\rm h} ^{1/3}  m _ {\rm s} ^{2/3} \ll |  \vec k |  \ll k _{\rm max} ) 
  \label{again} 
\end{eqnarray} 
Thus, in this case  the effect of soft gluons  on the growth rate is negligible. 
Note that the range in which  Eq.~(\ref{again}) holds only exists
when $ v _ {\rm h} ^ z \ll ( m _ {\rm h} / m _ {\rm s} ) ^{2/3} $.

\section{Plasma instabilities and bottom-up scenario}
\label{sc:bottomUp} 

We will now address the question whether and how plasma instabilities affect
the parametric estimates of the bottom-up scenario \cite{bottomUp}. 
It is clear that they have a dramatic 
impact if they can grow as large as in QED, i.e., if they do not saturate until 
$ A \sim p/g $ \cite{arnoldThermalization}. 
In this case the unstable modes grow so large that they change the 
momenta of the hard particles by an amount of order one within a time of order
$ m ^ {-1}$, which would lead to a quick isotropization.
We will  assume that
the instabilities already saturate when $ A   \sim k  /g $, as
indicated by recent lattice simulations \cite{arnoldFate,romatschkeFate}.
Then the effect of the instabilities on the hard gluons is much less dramatic. 

We only discuss the first stage of the bottom-up scenario when 
the hard and the soft contributions to 
$ \Pi  ^{{\mu  \nu  } } $ are of the same order of magnitude, i.e.,  
$ m _ {\rm s} \sim m _ {\rm h} \sim m $
and the distribution of soft gluons is nearly isotropic.
We have seen    in Sec.~\ref{sc:two} that there are
plasma instabilities irrespective of the numerical factors in $ m _ {\rm s} $
and  $ m _ {\rm h} $.  

The typical longitudinal momentum of hard gluons decreases due to the
expansion. If they were free streaming, one would have 
$ p _ {\rm h} ^ z \propto \tau  ^{-1} $.  
On the other hand, $ p _ {\rm h} ^ z$ increases due to multiple scattering.
Combining these two effects one obtains 
$ p _ {\rm h} ^ z \propto \tau  ^{-1/3} $ \cite{bottomUp}. 
We will now consider how the instabilities affect this estimate. 
  
In the linear regime the unstable modes cannot  be described as
particles, but rather as a coherent classical field. When its amplitude
becomes of order $  m   /g $ its equation of motion becomes non-linear. 
Due to the complicated interaction the
coherent nature of the gluon field will at least partly be lost. Parametrically
the occupation number of the $ |  \vec k |  \sim m $ modes is then 
$ f _ m \sim 1/g ^{2} $. 
A quantitative description of their dynamics by a Boltzmann
equation is 
not possible because it would require that $ f _ m \ll 1/g ^{2}  $. 
But since this is  right at the
border of validity of the Boltzmann equation one should still be able
to use it for parametric estimates which is what we will do in the following.

Consider  $ 2 \to 2 $ scattering of a hard
gluon with a  $  |  \vec k  | \sim m $ gluon  with a momentum
transfer of order $ m $.  
The corresponding cross section $ \sigma  $ is of order  $ g ^ 4 /m ^ 2
$. Therefore the  rate at which a
hard gluon experiences such collisions is 
\begin{eqnarray}
  \frac{ d N _{\rm col} }{d \tau } \sim \sigma   n _ m f _ m \sim m 
\end{eqnarray} 
where we have used $ n _ m \sim m ^ 3 /g ^{2} $ for the number density of 
$ |  \vec k |  \sim m $ gluons and 
where  $ f _ m $ entered as a Bose enhancement factor. 
The hard gluons experience many random collisions each with  
a momentum transfer of order $ m $  so that 
\begin{eqnarray}
  ( p  _ {\rm h} ^ z ) ^ 2 \sim N _{\rm col} m  ^ 2 
  \sim \tau  \frac{ d N _{\rm col} }{d \tau } m  ^ 2
\end{eqnarray} 
In the bottom-up scenario $ m ^ 2  \sim g ^ 2 n _ {\rm h} /Q \sim Q ^ 2/(Q \tau  ) $
where  
$ n _ {\rm h} $ is the number density of hard gluons. Thus these collisions lead to
\begin{eqnarray}
  p _ {\rm h} ^ z \sim Q ( Q \tau  ) ^{-1/4} 
  \label{pz} 
\end{eqnarray} 
which is larger than in the bottom-up scenario where 
$ p _ {\rm h} ^ z \sim Q ( Q \tau  ) ^{-1/3}  $. 

We have thus  seen that plasma instabilities indeed lead to a certain
isotropization  
in an expanding gluon plasma, even if they are saturated by non-abelian
interactions. But the isotropization is far from complete, and whether the
instabilities really lead to equilibration is not clear. 

Let us finally use Eq.~(\ref{pz}) to  determine $  k _{\rm max} $ of
Eq.~(\ref{kmax}), i.e., the largest  
$ |  \vec k |  $  for which there are plasma instabilities.  
We have
\begin{eqnarray}
  v _ {\rm h} ^ z \sim \frac{ p _ {\rm h} ^ z }{Q} \sim ( Q \tau  ) ^{-1/4}
\end{eqnarray} 
which gives
\begin{eqnarray}
  k _{\rm max} \sim \frac{ Q}{( Q \tau  ) ^{1/2}} ( Q \tau  ) ^{1/4}  \sim 
  Q ( Q \tau  ) ^{-1/4} 
  \label{new.2} 
\end{eqnarray} 
It is interesting that the result (\ref{new.2}) coincides with 
$ p _ {\rm h} ^ z $ in Eq.~(\ref{pz}).

\vspace{.5cm}
{\bf Acknowledgments} 
I would like to thank R.~Baier, M.~Laine, G.~D.~Moore, A.~H.~Mueller, P.~Romatschke,
K.~Rummukainen, D.~Schiff,  
A.~Shoshi,  and L.~Yaffe
 for interesting discussions or comments.
This work was supported in part by the DFG under grant FOR 339/2-1, and through the
DFG funded Graduate School GRK 881.



         


\begin{thebibliography}{99}

\bibitem{saturation}
See Ref.~\cite{bottomUp} and references therein. 

\bibitem{muellerBoltzmann}
A.~H.~Mueller,
  ``Toward equilibration in the early stages after a high energy heavy ion
  collision,''
  Nucl.\ Phys.\ B {\bf 572} (2000) 227
  [arXiv:hep-ph/9906322];
``The Boltzmann equation for gluons at early times after a heavy ion
collision,''
Phys.\ Lett.\ B {\bf 475} (2000) 220
[arXiv:hep-ph/9909388].

\bibitem{raju}
J.~Bjoraker and R.~Venugopalan,
``From colored glass condensate to gluon plasma: Equilibration in high  energy
heavy ion collisions,''
Phys.\ Rev.\ C {\bf 63} (2001) 024609
[arXiv:hep-ph/0008294].

\bibitem{wong}
S.~M.~H.~Wong,
  ``Thermal and chemical equilibration in a gluon plasma,''
  Nucl.\ Phys.\ A {\bf 607}, 442 (1996)
  [arXiv:hep-ph/9606305].

\bibitem{bottomUp}
R.~Baier, A.~H.~Mueller, D.~Schiff and D.~T.~Son,
``'Bottom-up' thermalization in heavy ion collisions,''
Phys.\ Lett.\ B {\bf 502} (2001) 51
[arXiv:hep-ph/0009237].

\bibitem{mrowczynskiInitial}
S.~Mrowczynski,
``Plasma instability at the initial stage of ultrarelativistic heavy ion
collisions,''
Phys.\ Lett.\ B {\bf 314} (1993) 118.

\bibitem{arnoldInstabilities} P.~Arnold, J.~Lenaghan and G.~D.~Moore,
``QCD plasma instabilities and bottom-up thermalization,''
JHEP {\bf 0308} (2003) 002
[arXiv:hep-ph/0307325].


\bibitem{arnoldAbelianization}
P.~Arnold and J.~Lenaghan,
``The abelianization of QCD plasma instabilities,''
Phys.\ Rev.\ D {\bf 70} (2004) 114007
[arXiv:hep-ph/0408052].

\bibitem{arnoldThermalization}
P.~Arnold, J.~Lenaghan, G.~D.~Moore and L.~G.~Yaffe,
``Apparent thermalization due to plasma instabilities in quark gluon plasma,''
Phys.\ Rev.\ Lett.\  {\bf 94} (2005) 072302
[arXiv:nucl-th/0409068].

\bibitem{hydrodynamics}
U.~W.~Heinz,
``Thermalization at RHIC,''
AIP Conf.\ Proc.\  {\bf 739} (2005) 163
[arXiv:nucl-th/0407067].

\bibitem{romatschke2d}
A.~Rebhan, P.~Romatschke and M.~Strickland,
``Hard-loop dynamics of non-Abelian plasma instabilities,''
Phys.\ Rev.\ Lett.\  {\bf 94} (2005) 102303
[arXiv:hep-ph/0412016].

\bibitem{dumitru}
A.~Dumitru and Y.~Nara,
``QCD plasma instabilities and isotropization,''
Phys.\ Lett.\ B {\bf 621} (2005) 89
[arXiv:hep-ph/0503121].

\bibitem{arnoldFate}
P.~Arnold, G.~D.~Moore and L.~G.~Yaffe,
``The fate of non-abelian plasma instabilities in 3+1 dimensions,''
arXiv:hep-ph/0505212.


\bibitem{romatschkeFate}
A.~Rebhan, P.~Romatschke and M.~Strickland,
``Dynamics of quark-gluon plasma instabilities in discretized hard-loop
approximation,''
arXiv:hep-ph/0505261.

\bibitem{muellerModified}
A.~H.~Mueller, A.~I.~Shoshi and S.~M.~H.~Wong,
``A possible modified 'bottom-up' thermalization in heavy ion collisions,''
arXiv:hep-ph/0505164.

\bibitem{heinz}
U.~W.~Heinz,
``Kinetic Theory For Nonabelian Plasmas,''
Phys.\ Rev.\ Lett.\  {\bf 51} (1983) 351.

\bibitem{vlasovQED}
For a review of the Vlasov equations in QED see, e.g., Ref.~\cite{landau10}. 

\bibitem{landau10}
E.M.~Lifshitz, L.P.~Pitaevskii, {\em Physical Kinetics} (Pergamon Press,
Oxford 1981).

\bibitem{romatschkePi}
P.~Romatschke and M.~Strickland,
``Collective modes of an anisotropic quark gluon plasma,''
Phys.\ Rev.\ D {\bf 68} (2003) 036004
[arXiv:hep-ph/0304092].

\bibitem{kraemmer}
For a review see,  e.g., 
U.~Kraemmer and A.~Rebhan,
``Advances in perturbative thermal field theory,''
Rept.\ Prog.\ Phys.\  {\bf 67} (2004) 351
[arXiv:hep-ph/0310337].

\end{thebibliography}
\end{document}